\documentclass[11pt,a4paper]{article}
\usepackage[T1]{fontenc}
\usepackage[utf8]{inputenc}
\usepackage[english]{babel}
\usepackage[margin=24mm]{geometry}
\usepackage{amsmath,amssymb}
\usepackage{newtxtext,newtxmath}
\usepackage{bm}
\usepackage{microtype}
\usepackage{graphicx}
\usepackage{caption}
\usepackage[numbers,sort&compress]{natbib}
\usepackage{xurl}
\usepackage[hidelinks]{hyperref}
\usepackage{bookmark}
\usepackage{needspace}
\graphicspath{{figures/}}
\renewcommand{\eqref}[1]{\textup{(\ref{#1})}}

\providecommand{\pa}{\phi^{(a)}}
\providecommand{\pb}{\phi^{(b)}}
\providecommand{\Tab}{T^{(a,b)}}
\providecommand{\Tba}{T^{(b,a)}}

\providecommand{\galign}{g_{\mathrm{align}}}
\providecommand{\Jint}{J_{\mathrm{int}}}

\title{\bfseries Bearing, heading and the perceptual geometry of collective motion}
\author{Mohammad Salahshour\thanks{Correspondence: \href{mailto:salahshour.mohammad@gmail.com}{salahshour.mohammad@gmail.com}}\\[0.5em]
\small Department of Collective Behaviour, Max Planck Institute of Animal Behavior,\\
\small Konstanz, Germany\\[0.3em]
\small Centre for the Advanced Study of Collective Behaviour, University of Konstanz,\\
\small Konstanz, Germany\\[0.3em]
\small Department of Biology, University of Konstanz, Konstanz, Germany}
\date{}

\hypersetup{pdftitle={Bearing, heading and the perceptual geometry of collective motion},pdfauthor={Mohammad Salahshour}}
\begin{document}
\maketitle

\begin{abstract}
Traditional models of collective motion commonly assume that animals turn to match their neighbours' directions of travel. In the harmonic theory, decisions follow from a neural representation of direction, and a neighbour enters only as a stimulus at a bearing, the direction in which it lies. Here we show that a neighbour's spatial extent allows its orientation to enter through projection geometry alone, and the theory generates a matrix of coupling coefficients indexed by the neighbour's bearing and aspect. Responses to bearing alone form one edge of this matrix, and classical heading matching is a special case on its diagonal, one ray through a two-parameter family. Independently, an animal that responds only to neighbour position, perceived with a delay, acquires an alignment coefficient set by the slope of its distance-weighted attraction profile, so that it can align, anti-align or reverse sign. Alignment need not be a rule to be measured. In schooling fish and walking pigeons with body orientation measured independently of travel, social responses are recovered, including bearing responses in the fish. Heading excess depends on how turning is measured, appearing in the fish with forward but not centred responses and remaining unresolved in pigeons. Controls show why neither outcome identifies a direct alignment rule. In fish recordings with motion-assigned body polarity, heading-dependent structure departs from Vicsek form and fails the coefficient test for a common angular potential. Whether animals align thereby becomes a measurable question, with geometric, perceptual and behavioural origins that can be distinguished.
\end{abstract}


\section*{Introduction}
For half a century, the movement of animal groups has been modelled by assuming that individuals turn to match the directions of travel of their neighbours \cite{Aoki1982,Reynolds1987,Vicsek1995,Couzin2002}. In the Vicsek model it is the only interaction \cite{Vicsek1995,Chate2008,Vicsek2012}, in the hydrodynamic theories built on it the alignment term supplies the symmetry that makes long-range order possible in two dimensions \cite{Toner1995,Toner1998,Marchetti2013}, and in zonal models it is one of three concentric rules \cite{Couzin2002,Couzin2005}. Whatever else these models differ on, they assume that a moving animal steers towards its neighbours' directions of travel.

The evidence comes from inferring interaction rules directly from tracked trajectories \cite{Lukeman2010,Katz2011,Herbert-Read2011,Gautrais2012,Calovi2018,Escobedo2020,Heras2019}, by regressing an individual's turning onto its neighbours' relative positions and velocities. Two directions matter here: an animal's heading, the direction its body faces, and its direction of travel. In almost all such datasets the heading is not measured but reconstructed from the displacement of the centroid between frames, so the two are identical by construction and any response to a neighbour's motion is interpreted as a response to its heading.

The heading-matching assumption has not gone unchallenged. Golden shiners show no evidence of explicit matching of body orientation \cite{Katz2011}, and mosquitofish turn in response to a neighbour's position with only weak evidence of a response to its orientation \cite{Herbert-Read2011}. Marching locusts are captured by a ring-attractor model that represents each conspecific solely by its bearing, the direction in which it lies \cite{Sayin2025}, and that orders globally only when those bearings are held in an allocentric frame \cite{Salahshour2025Allocentric}. Models without access to neighbours' headings also produce collective order \cite{Bastien2020,Barberis2016}. Collective order establishes neither the presence nor the necessity of a heading-matching rule.

We recently derived a framework in which movement decisions follow from coarse-grained neural dynamics on a ring-shaped internal representation of direction \cite{Salahshour2025Harmonic}, of the kind characterized across insects and vertebrates \cite{Zhang1996,Taube2007,Seelig2015,Kim2017,TurnerEvans2017,Green2017,Hulse2021,Wilson2023}. We call it the harmonic theory (Supplementary Information, \ref*{SI-SI:summary}). For a neighbour seen as a point, every social term in it couples an individual's heading $\pa$ to the bearing $\Tab$ of that neighbour, and no term couples $\pa$ to the neighbour's own heading $\pb$, because a stimulus on the sensory ring has a location before it has an orientation. We call this form the point-stimulus theory. Within it, whatever alignment a group displays is generated by the geometry of relative motion and is nowhere encoded as a rule.

At first sight the test is obvious: measure body orientation independently of the direction of travel and decompose the observed turning into a component driven by where neighbours are and one driven by which way they face. It is harder than it looks. The two channels can be separated only where they predict different turns. In a polarized group the bearing and heading predictors become collinear, and in a disordered group they separate but there is little signal to attribute. Identifiability and estimability pull in opposite directions.

What makes the question tractable is a result that at first seems to make it worse. Suppose an individual responds only to neighbour position but perceives that position with a finite delay. A lagged pursuer steers toward where its neighbour used to be. Depending on how its attraction falls off with distance, it appears to anti-align, to align, or to switch from one to the other with range. Measuring alignment therefore cannot falsify a bearing-only theory. We show that the discriminating content lies one level down. A delay adds two corrections to the bearing response, one following the change in a neighbour's distance and one following the drift of its bearing, and the measured attraction profile and its slope fix their ratio, a constraint we call the coefficient lock. A genuine alignment rule has its own gain and no reason to obey it.

Here we show that the harmonic theory contains the bear\-ing--heading test as soon as a neighbour is allowed to have a body. An extended neighbour covers an angular interval on the sensory ring whose width and midpoint depend on its orientation, so orientation enters through projection geometry alone and every coefficient gains a second index. The point-stimulus theory is one edge of the resulting matrix, and classical heading matching is a special case on its diagonal, one ray through a two-parameter family. We fit bearing and heading as separate blocks and apply the fit to public recordings of schooling fish and walking pigeons whose body orientation is measured independently of travel. The heading excess is the turning variance that neighbour headings explain beyond their bearings, minus the same quantity in a null in which each animal is given neighbours from another trial. Bearing responses are recovered in the fish, as is a social response in small pigeon groups. Heading excess is resolved in the fish with forward but not centred turning responses, and remains unresolved in pigeons. Controls with known interactions show why these outcomes do not identify a direct alignment mechanism. In fish recordings whose front--back polarity is assigned from motion, heading-dependent structure is not of Vicsek form and fails the coefficient test for a common angular potential in the tracked coordinates. An alignment coefficient is thereby turned from an assumed rule into a quantity whose geometric, perceptual and behavioural origins can be distinguished.

\section*{Results}

\subsection*{A neighbour with a body: bearing and heading in one matrix}
\label{ResultsTheory}
\begin{figure}[tbp]
\centering
\includegraphics[width=\textwidth]{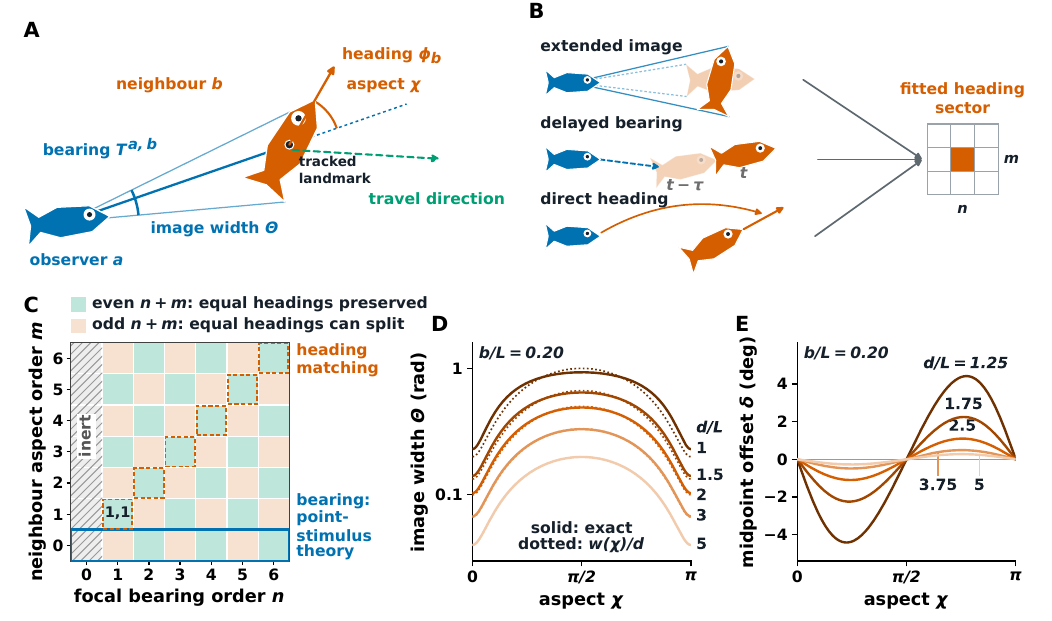}
\caption{\textbf{Body geometry connects bearing and heading, and one fitted response has several origins.} (\textit{A}) An extended neighbour $b$ casts an angular image on observer $a$. Bearing $T^{a,b}$ locates its tracked landmark, $\phi_b$ is its heading, $\chi$ its aspect to the line of sight and $\Theta$ its image width; heading can differ from the direction of travel. (\textit{B}) An extended image, a delayed bearing response and a direct heading response can each contribute to the same fitted heading sector, the $(1,1)$ cell of \textit{C}. (\textit{C}) The coefficient matrix of Eq.~\eqref{eq:mttwoindex}, drawn as a grid of cells. The blue edge $m=0$ is the point-stimulus theory; the dashed diagonal $n=m$ holds the pure heading-difference interactions ($K_{nn}=L_{nn}$), with Vicsek alignment at $n=1$. Cells with even $n+m$ keep equal headings equal and odd cells can split them; the hatched $n=0$ column exerts no torque. (\textit{D}) Exact angular width $\Theta$ of an ellipse of length $L$ and width $b$ ($b/L=0.20$) against aspect $\chi$ at five ranges $d$ (solid), and its far-field form $w/d$, where $w$ is the projected width (dotted); $\chi=0$ is tail-on, $\pi/2$ broadside and $\pi$ head-on. (\textit{E}) The image midpoint of the same ellipse shifts with aspect and changes sign across broadside, at the five ranges labelled in that panel; this phase generates the sine family of Eq.~\eqref{eq:mttwoindex}. Spectra, shapes and landmark effects: \ref*{SI-SI:subtense}--\ref*{SI-SI:gauge}, Fig.~\ref*{SI-fig:pnas:geometry}; parity rule and curl: \ref*{SI-SI:parity}, Fig.~\ref*{SI-EDFig2}.}
\label{Fig1}\label{FigOverview}\label{FigGeometry}
\end{figure}

An animal can respond to a neighbour's orientation without ever measuring it. A neighbour's body subtends an angle at the observer's eye, and that angle changes as the neighbour turns (Fig.~\ref{Fig1}\textit{A}); this image is all the harmonic theory needs. It is also only one of three routes by which a heading-shaped term can enter a fitted interaction (Fig.~\ref{Fig1}\textit{B}); the other two, perceptual delay and a direct response to heading, are treated below. Let animal $a$ have heading $\pa$ and perceive animal $b$ at bearing $\Tab$ and distance $d$. For a point stimulus the theory's decision landscape is $H^{(a\leftarrow b)}=-\sum_{n\ge1}\Jint(d)\,c_nM_n\cos(n\psi)$ with $\psi=\pa-\Tab$, a sum over harmonics of the bearing error weighted by a distance-dependent interaction strength $\Jint(d)$, a sensory filter $c_n$ and a neural decision filter $M_n$ \cite{Salahshour2025Harmonic}. Each animal turns down its own landscape, $\Gamma^{(a)}=-\eta\,\partial_{\pa}H^{(a\leftarrow b)}$, with angular mobility $\eta$ (\ref*{SI-SI:summary}, \ref*{SI-SI:prescriptions}).

A body of length $L$ and width $b$ (a length, distinct from the neighbour label $b$) seen at aspect $\chi=\pb-\Tab$ subtends an angle $\Theta(\chi,d)$ that is widest broadside and narrowest end-on (Materials and Methods). For a silhouette of constant angular brightness whose midpoint is offset by $\delta$ from the tracked landmark, integrating the point response over the image gives
\begin{align}
 H^{(a\leftarrow b)}&=-\sum_{n\ge1}Q_n\cos[n(\psi-\delta)],\nonumber\\
 Q_n&=\Jint c_nM_n\,\Theta\,\operatorname{sinc}(n\Theta/2),
 \label{eq:mtimage}
\end{align}
with $\operatorname{sinc}(x)=\sin(x)/x$, exact for an image of any width. The body therefore adds one amplitude factor, $\Theta\operatorname{sinc}(n\Theta/2)=(\pi/2)S_n(\Theta)$ with object filter $S_n(\Theta)=4\sin(n\Theta/2)/(n\pi)$, and one phase, the midpoint offset $\delta$. Both are properties of the animal being looked at rather than of the one looking (\ref*{SI-SI:cumulant}--\ref*{SI-SI:objectfilter}).

Because $\Theta$ and $\delta$ depend on the aspect, and the aspect contains the neighbour's heading, resolving them into aspect harmonics gives
\begin{align}
H^{(a\leftarrow b)}=-\sum_{n,m}\big[&K_{n,m}(d)\cos(n\psi)\cos(m\chi)\nonumber\\
 &+L_{n,m}(d)\sin(n\psi)\sin(m\chi)\big].
\label{eq:mttwoindex}
\end{align}
The index $n$ counts harmonics of the perceiver's bearing error and $m$ harmonics of the neighbour's aspect. We call each coefficient pair $(K_{n,m},L_{n,m})$ a cell of the matrix and the cosine and sine sets its two families; the bearing cells ($m=0$) and the heading cells ($m\ge1$) are the two sectors that the regression below fits as separate blocks. The image width populates the cosine family, the midpoint shift populates the sine family, and bilateral symmetry excludes the two mixed products (\ref*{SI-SI:aspectprofile}, \ref*{SI-SI:centro}). The point-stimulus theory is the edge $m=0$ of this matrix (Fig.~\ref{Fig1}\textit{C}). A pure heading-difference interaction of order $n$ is the diagonal cell $(n,n)$ with $K_{n,n}=L_{n,n}$, because $\cos n(\pa-\pb)=\cos n\psi\cos n\chi+\sin n\psi\sin n\chi$. The $(1,1)$ cell alone contains an alignment term with weight $(K_{1,1}+L_{1,1})/2$ and a second term with weight $(K_{1,1}-L_{1,1})/2$ that locks the sum of the two headings to the line of sight. Vicsek alignment, the first-harmonic heading matching of the classical models, is one ray through this two-parameter family (\ref*{SI-SI:sector11}).

Geometry determines which cells can be occupied. Turning the neighbour changes its image width (Fig.~\ref{Fig1}\textit{D}), and for an outline that is centrally symmetric about the tracked landmark the width and the midpoint repeat every half turn, so only even aspect orders $m$ appear, exactly and at every range, in both families. Its image midpoint still moves as the body turns, so the sine family is populated even for a centred ellipse (Fig.~\ref{Fig1}\textit{E}). Fore--aft asymmetry admits odd orders through perspective alone, and so does the choice of tracking landmark: shifting the landmark along the body axis by a distance $s$ turns a point-image response into the odd-$m$ terms $K_{n,1}\simeq-sK'_{n,0}(d)$ and $L_{n,1}\simeq-n(s/d)K_{n,0}(d)$, so odd-$m$ coefficients must be interpreted relative to a specified landmark (\ref*{SI-SI:centro}--\ref*{SI-SI:gauge}; Fig.~\ref*{SI-fig:pnas:geometry}). For small images the phase-free cosine family separates into a perceiver factor that sets the $n$-spectrum and a neighbour factor that sets the $m$-spectrum, which predicts that manipulating the perceiver and manipulating the neighbour change different spectra (\ref*{SI-SI:factorization}).

\subsection*{Which interactions keep a pair aligned, and when a common potential exists}
\label{ResultsParity}
Every pairwise interaction is used twice, by $a$ looking at $b$ and by $b$ looking at $a$, and exchanging the two animals reverses their common line of sight by $\pi$. A harmonic cell then acquires the sign $(-1)^{n+m}$ while its indices are exchanged (Materials and Methods). For identical animals, cells with even $n+m$ exert no torque that separates two equal headings, a consensus state, whereas cells with odd $n+m$ generally split them (Fig.~\ref{Fig1}\textit{C}). The alignment diagonal is always even, which is why heading matching at any order preserves consensus (\ref*{SI-SI:headingparity}). Whether nearby headings converge is a separate stability question, and a many-animal group need not order (\ref*{SI-SI:exchange}--\ref*{SI-SI:consensus}, \ref*{SI-SI:attraction}; Fig.~\ref*{SI-fig:si:consensusstability}).

A stronger question is whether both animals descend one common angular potential at fixed positions, as they do in the point-stimulus theory~\cite{Salahshour2025Harmonic}. For the periodic fields considered here this holds if and only if the torque field has zero curl in the two headings \cite{CrossHohenberg1993,Lee2012SmoothManifolds}, which gives the coefficient conditions
\begin{align}
nm\,[K_{n,m}-(-1)^{n+m}K_{m,n}]&=0,\nonumber\\
nm\,[L_{n,m}-(-1)^{n+m}L_{m,n}]&=0
\label{eq:mtintegrability}
\end{align}
for every pair of cells (\ref*{SI-SI:integrability}). Equation~\eqref{eq:mtintegrability} sorts the matrix into three parts. Cells that satisfy both brackets are reciprocal (exchange-symmetric), meaning that exchanging the animals leaves the directed interaction landscape unchanged, and variational, meaning that both torques derive from one potential. A mismatch on the bearing edge is non-reciprocal yet remains variational, because $nm=0$ removes it from the curl. In the chosen landscape convention, the point-stimulus theory is exchange-asymmetric but still potential-based (\ref*{SI-SI:integrability}). A mismatch in the interior produces curl and rules out a common potential, and the leading image approximation already produces such a mismatch (Fig.~\ref*{SI-EDFig2}\textit{c,d}; \ref*{SI-SI:integrability}). Self-propelled motion changes the positions and can increase that potential, so milling or a rotating pair does not by itself demonstrate curl (\ref*{SI-SI:integrability}).

\subsection*{Perceptual delay generates alignment}
\label{ResultsDelay}
\begin{figure}[tbp]
\centering
\includegraphics[width=\textwidth]{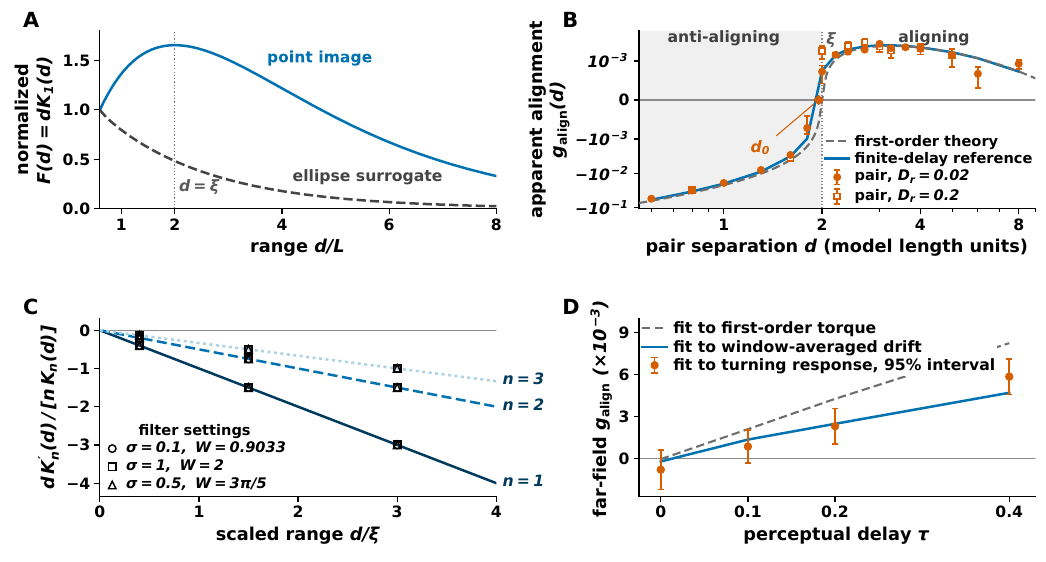}
\caption{\textbf{Perceptual delay predicts a constrained response whose sign depends on the image.} (\textit{A}) $F(d)=dK_1(d)$ for a point image and for a phase-free, aspect-averaged ellipse surrogate with $b/L=0.2$, constant angular brightness and envelope $e^{-d/\xi}$, $\xi/L=2$, each normalized at $d/L=0.6$. The maximum of the point curve predicts a sign reversal; the surrogate decreases throughout. (\textit{B}) Controlled pairs with the exponential point kernel, on a symmetric-log axis (shading, the anti-aligning side of $\xi$): first-order theory (dashed), a finite-delay reference on prescribed histories (solid) and pair estimates at two rotational-noise levels $D_r$ (markers, bars of one standard error, SE) share the sign profile. The dotted line marks $d=\xi$ and the diamond the fitted zero $d_0$ of the $D_r=0.02$ pairs, whose 95\% bootstrap interval is narrower than the marker. (\textit{C}) The radial--transverse ratio of Eq.~\eqref{eq:mtlock}, $-d/(n\xi)$ for this kernel (lines, $n=1$ to $3$), is the same for three filter settings (markers; $\sigma$ is the sensory width and $W$ the decision-bump width). (\textit{D}) Far-field apparent-alignment coefficient in moving five-agent groups against delay: the fit that includes the mixed angular terms the delay generates, applied to the simulated turning responses (markers, replicate-based 95\% intervals), to the window-averaged drift (solid) and to the first-order torque (dashed), 192 groups per delay. Details: \ref*{SI-SI:neighbourmotion}--\ref*{SI-SI:latencylock} (\textit{A,C}) and \ref*{SI-SI:signreversal}--\ref*{SI-SI:test3} (\textit{B,D}); Figs.~\ref*{SI-fig:si:extendeddelay}, \ref*{SI-EDFig3}, \ref*{SI-fig:pnas:delaydetail}.}
\label{Fig2}
\end{figure}

A neighbour's orientation enters the matrix through its image. A second route needs no image at all. If an animal acts on an estimate of neighbour position that lags the true position by a time $\tau$, its turning depends on how the neighbour moved during the delay, and to first order the bearing response acquires two corrections, one proportional to the rate at which the range changes and one to the rate at which the bearing drifts (Materials and Methods, Eq.~\eqref{eq:mtmaster}; \ref*{SI-SI:sensorydelay}--\ref*{SI-SI:differentiate}). For motion along body heading, averaging over the neighbour's bearing leaves a pure heading-difference torque, $-\galign\sin(\pa-\pb)$, whose strength for a neighbour moving at speed $v_b$ is
\begin{equation}
\galign(d)=-\frac{\eta\tau v_b}{2d}\,\frac{\mathrm d}{\mathrm dd}\big[dK_1(d)\big].
\label{eq:mtgalign}
\end{equation}
With sideslip, the same torque instead involves the neighbour's travel direction (\ref*{SI-SI:neighbourmotion}). Positive $\galign$ aligns and negative $\galign$ anti-aligns, and the sign is fixed by the measured bearing profile $K_1(d)$. Where $dK_1(d)$ increases the delay predicts anti-alignment, where it decreases it predicts alignment, and its stationary points are the zeros (\ref*{SI-SI:neighbourmotion}). For an exponential point kernel the single stationary point is the interaction range $\xi$, so a delayed pursuer anti-aligns inside $\xi$ and aligns beyond it (Fig.~\ref{Fig2}\textit{A,B}). The phase-free, aspect-averaged ellipse surrogate has no stationary point over the tested range, so whether a reversal occurs is set by the image model and the radial profile; this calculation does not cover the full phase-aware image (Fig.~\ref{Fig2}\textit{A}; \ref*{SI-SI:neighbourmotion}, Fig.~\ref*{SI-fig:si:extendeddelay}). The same delay also changes the stability of direct approach to a stationary target, a prediction that needs no neighbour (\ref*{SI-SI:bifurcation}).

The same delay constrains the two corrections jointly. Their radial and transverse amplitudes obey
\begin{equation}
\frac{\Gamma_n^{\mathrm{rad}}}{\Gamma_n^{\mathrm{trans}}}
=\frac{dK'_n(d)}{nK_n(d)},\qquad K_n(d)\ne0,
\label{eq:mtlock}
\end{equation}
from which the delay, the speed and every separation-independent gain cancel pointwise; we call this relation the coefficient lock (Fig.~\ref{Fig2}\textit{C}; \ref*{SI-SI:latencylock}). A direct heading rule has an independent gain and is not constrained by this relation. A general optic-flow response is distinguishable when it violates the lock and locally indistinguishable at first order when it obeys it (\ref*{SI-SI:opticflow}).

Simulations recover the predicted sign profile. Controlled pairs started at prescribed separations resolve negative near-field and positive far-field coefficients, with the fitted zero at $d_0=1.951$ (95\% interval $1.921$ to $1.983$), just inside the first-order value $\xi=2$ because a finite delay shifts the zero slightly inward (Fig.~\ref{Fig2}\textit{B}; \ref*{SI-SI:signreversal}, \ref*{SI-SI:latency:pairrecovery}; Fig.~\ref*{SI-fig:pnas:delaydetail}b). In freely moving five-agent groups the far-field coefficient is recovered when the fit includes the mixed angular terms that the delay generates, and not otherwise (\ref*{SI-SI:test3}); with those terms it is resolved at $\tau=0.2$ and $0.4$ and not at $\tau=0.1$, and the same fit applied to the noise-free drift and to the first-order torque gives the two reference curves (Fig.~\ref{Fig2}\textit{D}).

\subsection*{A neighbour's image changes the collective state, without a heading rule}
\label{ResultsMotion}
\begin{figure}[tbp]
\centering
\includegraphics[width=\textwidth]{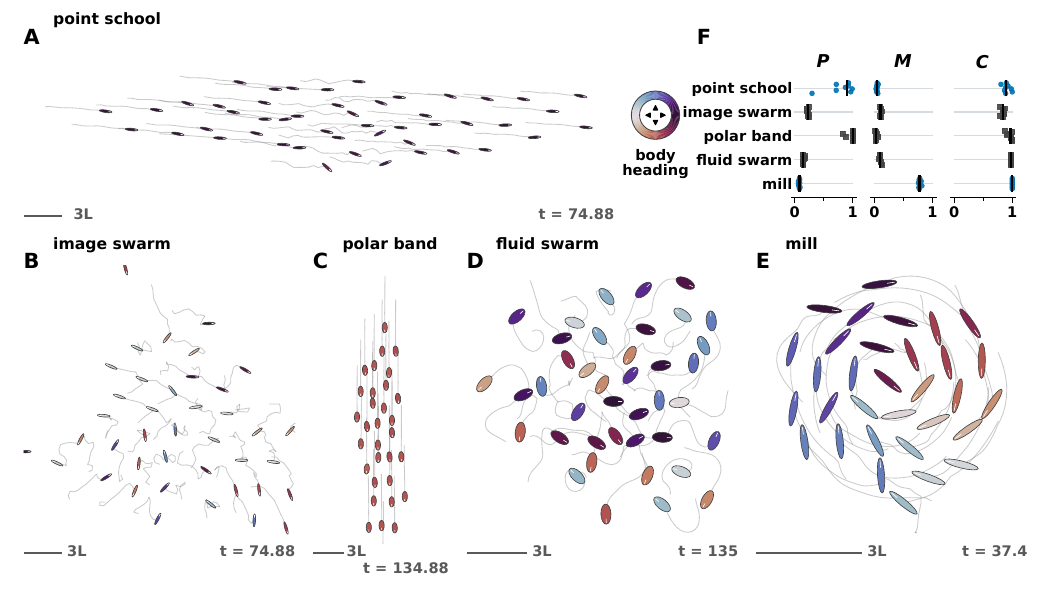}
\caption{\textbf{A neighbour's image changes the collective state that bearing responses produce, with no heading rule.} (\textit{A,B}) One matched comparison: the same positions, headings and noise, with the neighbour seen as a point (\textit{A}) or as the exact image of a centred ellipse (\textit{B}). The point-bearing response forms a polarized school and the image a swarm with low heading order, shown at the same model time $t$ and on the same scale (Movie~S4). (\textit{C}) At another setting the full image forms a persistent polar band (Movie~S5). (\textit{D}) In unbounded space it sustains a cohesive fluid swarm (Movie~S6). (\textit{E}) The first bearing harmonic alone, without a wall or higher harmonics, produces a cohesive mill (Movie~S9). Each snapshot is taken three-quarters of the way through its run, at the model time printed in the panel, and framed on the largest cluster (agents linked within 2.6 body lengths; agents outside the frame are cut off); bodies are coloured by heading on the cyclic key, outlined, with a white tick at the front; grey trails span two model-time units; scale bars show three body lengths. (\textit{F}) For eight runs of each setting, the mean over the second half of the run of the polarization $P$ (length of the mean heading vector), the normalized angular momentum $M$ about the group centre (high in a mill) and the largest-cluster fraction $C$, with medians as black ticks and dots offset vertically only to separate them; circles are point-bearing settings on the $m=0$ edge of Fig.~\ref{Fig1}\textit{C} (point school $n=1$ to $3$, mill $n=1$) and squares exact-image settings, which add the even-$m$ cells of the image (image swarm and polar band $n=1$ to $3$, fluid swarm $n=1$). The settings differ in population size, noise, image parameters, speed, gains, interaction range, filter widths, collision parameters and boundary (\ref*{SI-SI:motion_atlas}); none includes a prescribed heading term. Equations, parameters and further patterns: \ref*{SI-SI:motion_atlas}, Fig.~\ref*{SI-fig:si:motioncontrols}.}
\label{FigMotion}
\end{figure}

The cells describe interactions, not group patterns. Groups driven by point-stimulus bearing responses are known to school, band and aggregate \cite{Salahshour2025Harmonic}, so we simulated self-propelled groups to ask what the neighbour's image adds (Fig.~\ref{FigMotion}; \ref*{SI-SI:simulator}, \ref*{SI-SI:motion_atlas}). In a matched comparison with identical initial positions, headings and noise, replacing the point stimulus by the exact image of a centred ellipse turns a polarized school into a swarm with much lower heading order (Fig.~\ref{FigMotion}\textit{A,B}; Movie~S4). The full image is no obstacle to order. At another parameter setting it forms a persistent polar band from random headings (Fig.~\ref{FigMotion}\textit{C}; Movie~S5; Fig.~\ref*{SI-fig:pnas:bandformation}), and in unbounded space it sustains a cohesive fluid swarm whose members keep exchanging neighbours (Fig.~\ref{FigMotion}\textit{D}; Movie~S6; Fig.~\ref*{SI-fig:si:fluidmotion}). The first bearing harmonic alone, without a wall or higher harmonics, produces a cohesive mill (Fig.~\ref{FigMotion}\textit{E}; Movie~S9), and eight simulations of each setting show these patterns of order and cohesion (Fig.~\ref{FigMotion}\textit{F}). None of these settings includes a prescribed heading-matching term; in the image settings the neighbour's heading enters only through its image. Adding an explicit heading-matching term aligns a group whether or not bearing attraction is present, but the group stays together only when attraction is present, so heading order and cohesion are distinct outcomes (Movie~S10; Fig.~\ref*{SI-fig:si:motioncontrols}). Movies~S1 and S2 compare the point and image kernels at matched parameters (Movie~S1 shows both original and matched interaction strengths), Movie~S3 isolates the image midpoint phase, and Movies~S7 and S8 show heading order from images averaged over bearing: a nematic state, in which parallel and antiparallel headings are equivalent, from an ellipse, and a polar state from a triangle of unit total intensity (\ref*{SI-SI:motion_atlas}, \ref*{SI-SI:videos}). These simulations show which behaviours the model can produce; determining whether animal groups use these interactions requires empirical tests.

\subsection*{What trajectory inference can and cannot see}
\label{ResultsIdentifiability}
\begin{figure}[tbp]
\centering
\includegraphics[width=\textwidth]{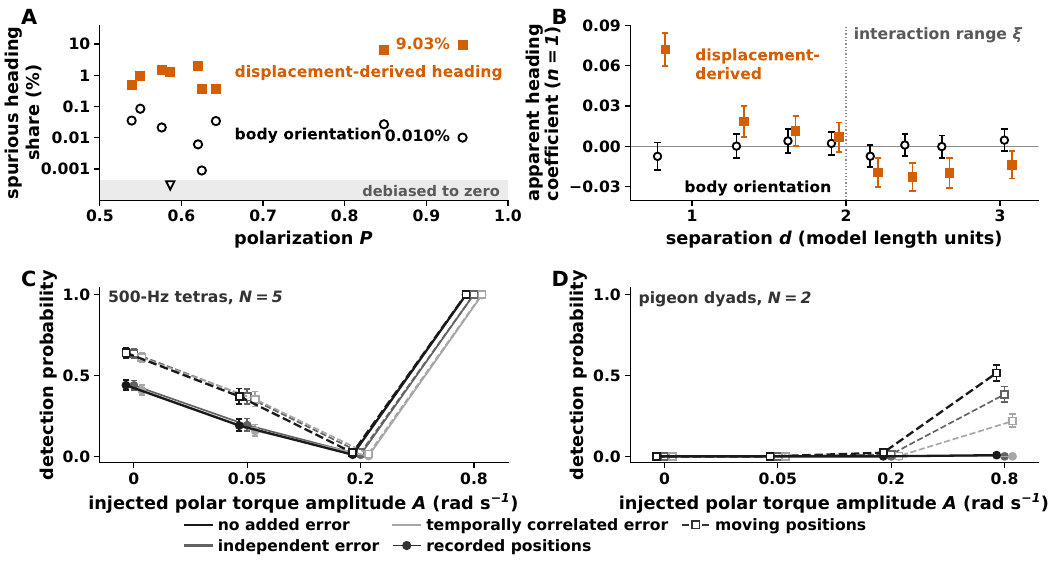}
\caption{\textbf{Observation can create an apparent heading signal or hide a known interaction.} (\textit{A}) Spurious heading share of the fitted bearing-plus-heading signal (noise-debiased, cross term excluded) in fits to identical bearing-only trajectories, using true body orientation (open circles) or displacement-derived heading (squares), against polarization. Each marker pair is one of nine simulated conditions at weak coupling ($h=-3$) differing in motor lag, sideslip and rotational noise. The grey band at the bottom indicates the estimate that debiases to exactly zero (triangle), and the printed values belong to the most polarized condition. (\textit{B}) With a motor lag and sideslip, and zero perceptual delay, the displacement channel (squares) yields an apparent heading coefficient that changes sign across the interaction range (dotted line), whereas body orientation (open circles) stays indistinguishable from zero. Bars are 1.96 standard errors clustered by 10-s blocks. (\textit{C}) Detection of a positive heading excess, with centred turning responses, when known polar torques $A\sin(\phi_b-\phi_a)$ per neighbour pass through the fish measurement procedure. Detection is not monotone in the amplitude $A$, and the zero-injection rate is not a false-positive rate, because bearing-only dynamics give a nonzero heading excess after observation and fitting. (\textit{D}) The corresponding detection test using the pigeon-dyad measurement procedure. In \textit{C} and \textit{D}, filled circles show recorded positions and open squares moving positions integrated from the simulated headings. Grey level gives the observation-error scenario (black, none added; mid grey, independent; light grey, temporally correlated). Amplitudes are equally spaced categories, and whiskers are binomial 95\% intervals over 1,000 repetitions at $A=0$ and 400 otherwise. All panels show simulation controls. In \textit{A} and \textit{B}, body orientation is the simulated angle $\phi$, whereas displacement-derived heading is the direction of net movement between successive recorded positions. Motor lag and sideslip allow movement to deviate from the body axis (simulation equations, \ref*{SI-SI:simulator}). Details for \textit{A} and \textit{B}: \ref*{SI-SI:test1b}, Figs.~\ref*{SI-EDFig5}--\ref*{SI-fig:si:observationchannel}. Full cohorts, nematic controls, recovery and coverage: \ref*{SI-SI:detectioncalibration}, Figs.~\ref*{SI-fig:si:detectioncalibration}--\ref*{SI-fig:si:headingcoverage}.}
\label{Fig3}\label{FigCalibration}
\end{figure}

Before analysing animal trajectories, we asked what the regression recovers when the interaction is known. The same trajectories can yield different heading signals depending on what is called heading. Fitting identical bearing-only trajectories with the true body orientation or with heading reconstructed from displacement gives a spurious heading share of the fitted bearing and heading signal of at most $0.085\%$ in the first case and up to $9\%$ in the second, across nine simulated conditions (Fig.~\ref{Fig3}\textit{A}; \ref*{SI-SI:test1b}). With a motor lag and sideslip between body axis and travel direction, the reconstructed channel even produces a heading profile that changes sign with distance, although the agents have neither a heading rule nor a perceptual delay (Fig.~\ref{Fig3}\textit{B}; \ref*{SI-SI:test1b}). Collective order adds a second difficulty. As headings become parallel the bearing and heading predictors become collinear: in a separate, strongly coupled bearing-only simulation fitted with true orientation the spurious share rises to $4\%$ at polarization $0.995$, and a perfect mill makes a heading predictor identical to a higher-order bearing predictor (\ref*{SI-SI:degeneracies}, \ref*{SI-SI:fom}, \ref*{SI-SI:test1}; Fig.~\ref*{SI-EDFig5}). The response span, the time interval over which turning is measured, adds a third. In a forward-difference control with known alignment, lengthening the span inflates the recovered alignment and attenuates the recovered bearing response, in opposite directions (\ref*{SI-SI:test2}).

We therefore asked directly whether the measurement procedures used below detect a known heading interaction. We added known polar torques, first-harmonic heading matching of the form $A\sin(\phi_b-\phi_a)$ per neighbour, to bearing-only dynamics fitted to and started from the recorded trajectories, and analysed the result exactly as we analysed the recordings, from preprocessing through the matched null, in which each focal animal is given neighbours from another trial or sequence, to clustered inference (Materials and Methods; \ref*{SI-SI:detectioncalibration}). With the centred response used here, fish detection is not monotone in the injected strength: it falls and then rises as the torque grows (Fig.~\ref{Fig3}\textit{C}). In the recorded-position control, an intermediate positive torque offsets the negative apparent polar coefficients of the bearing-only baseline (\ref*{SI-SI:detectioninterpretation}). With recorded positions the pigeon procedure misses even the strongest tested torque in almost every repetition, and letting positions evolve raises detection at that strength to roughly a fifth to a half of the repetitions, depending on the observation error (Fig.~\ref{Fig3}\textit{D}). Bearing-only generators, conversely, can yield a positive heading excess after observation and fitting. A heading excess is therefore neither a specific detector nor a monotone measure of heading matching, and a non-detection below cannot exclude it (\ref*{SI-SI:detectioncalibration}; Figs.~\ref*{SI-fig:si:detectioncalibration}--\ref*{SI-fig:si:headingcoverage}). With forward differences the strongest injected polar torques are detected more often in noise-free controls, but observation error can still bias their estimated strength, and frequent detection does not guarantee that the coefficient intervals cover the injected value (\ref*{SI-SI:detectioncalibration}, \ref*{SI-SI:filtertiming}). Fitting the distribution of neighbour positions, rather than turning rates, also encounters an identifiability limit: in a stationary, exchangeable group moving at constant speed, the statistic such a fit uses to detect polar attraction averages to zero even when attraction is strong (\ref*{SI-SI:occupancy}).

\subsection*{\phantomsection\label{ResultsPrimary}Fish and pigeons: bearing is recovered, and the heading excess depends on how turning is measured}
\begin{figure}[tbp]
\centering
\includegraphics[width=\textwidth]{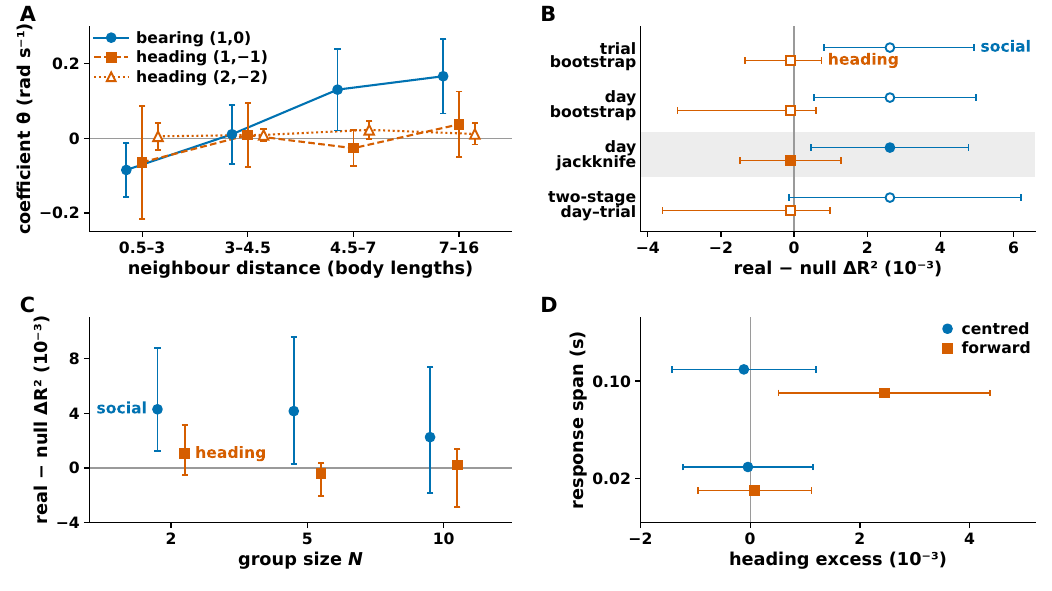}
\caption{\textbf{Bearing and social responses are recovered, while measured fish heading excess depends on response timing.} Panels \textit{A--C} use centred turning responses. (\textit{A}) Bearing $(1,0)$ and heading $(1,-1)$ polar and $(2,-2)$ nematic coefficients against neighbour distance in 500-Hz recordings of five rummy-nose tetras \cite{mckee2020,mchenry2020dryad}, 74 trials on ten recording days; bars are 95\% delete-one-day jackknife intervals with a $t_9$ reference. (\textit{B}) Paired real-minus-null excess in explained turning variance for the social block (all bearing and heading terms, blue) and the heading block (orange) under four resampling procedures (bars, 95\% intervals); the day jackknife, specified before testing, is highlighted. (\textit{C}) The same paired excesses in 3D-POP pigeons \cite{naik2023pop,naik2023edmond} against group size, with 95\% sequence-rematching intervals; for the pigeons the recovered response is the social block. (\textit{D}) Matched fish heading excess with centred (circles) and forward (squares) responses at two spans, using the original zero-phase filter. Each pair uses identical focal observations: $70{,}029$ at $0.10$~s and $71{,}929$ at $0.02$~s, from the same recordings as \textit{A,B}. Bars are pointwise $95\%$ ten-day jackknife intervals with a $t_9$ reference. At $0.10$~s the forward estimate is $2.44\times10^{-3}$ ($95\%$ interval $[0.52,4.37]\times10^{-3}$), and the paired forward-minus-centred increase is $2.56\times10^{-3}$ ($95\%$ interval $[0.31,4.80]\times10^{-3}$). Definitions, samples and procedures: \ref*{SI-SI:twosector}--\ref*{SI-SI:dayunit}, \ref*{SI-SI:pigeon:results}. Filtering sensitivity and the pigeon orientation-channel comparison are shown in Fig.~\ref*{SI-fig:si:empiricaltiming} (\ref*{SI-SI:empiricaltiming}).}
\label{FigMcKee}\label{Fig4}\label{FigChannels}\label{FigAnimals}
\end{figure}

We fitted the two sectors as separate blocks of a turning regression in distance bins, with arena terms, and compared each fit with its matched null (Fig.~\ref{FigMcKee}\textit{A--C}). The paired real-minus-null excess in explained turning variance is the tested quantity, with uncertainty resampled at the independent recording unit (Materials and Methods; \ref*{SI-SI:twosector}--\ref*{SI-SI:dayunit}). We label each fitted coefficient $\theta_{n,m}$ by its bearing harmonic $n$ and a signed aspect index $m$: $\theta_{n,m}$ multiplies the predictor formed by summing $-n\sin(n\psi+m\chi)$ over neighbours within each distance bin, with $\psi$ the bearing error and $\chi$ the aspect, so a negative $m$ selects the heading-difference combination of a cell and a positive $m$ its line-of-sight combination (Materials and Methods). Thus $(1,0)$ is the basic bearing response, $(1,-1)$ polar heading matching and $(2,-2)$ nematic matching, which treats a neighbour and its reverse alike.

In 500-Hz recordings of schooling rummy-nose tetras \cite{mckee2020,mchenry2020dryad}, heading is tracked from body landmarks and differs from the direction of travel by a circular spread of $19^\circ$ (\ref*{SI-SI:decoupling}--\ref*{SI-SI:tailbeat}). The fitted five-fish profile shows avoidance at close range and attraction at long range, with one of the two intermediate bins unresolved (Fig.~\ref{FigMcKee}\textit{A}); if these fish perceive with a delay, Eq.~\eqref{eq:mtgalign} applied to this profile predicts anti-alignment over every sampled interval, a conditional prediction rather than a detected reversal (\ref*{SI-SI:signreversal}, \ref*{SI-SI:twosector}). The combined social terms exceed their matched null by $2.6\times10^{-3}$ of the turning variance when whole recording days are left out in turn, with $p=0.022$, whereas resampling days and trials together gives an interval that includes zero (Fig.~\ref{FigMcKee}\textit{B}).

Whether a heading excess is resolved depends on how turning is measured. With the turning rate taken over a window centred on each observation, no heading excess is resolved under any tested resampling scheme, response span or polarization stratum, nor in fish whose lateral line, the flow-sensing organ, was ablated (Fig.~\ref{FigMcKee}\textit{B}; Fig.~\ref*{SI-EDFig6}; \ref*{SI-SI:matchednull}--\ref*{SI-SI:dayunit}). With the rate taken forward from each observation over $0.10$~s, the same observations give a heading excess of $2.4\times10^{-3}$ ($95\%$ interval $[0.5,4.4]\times10^{-3}$), and the paired difference between the two definitions is itself resolved; the $0.02$-s forward window remains unresolved, and the $0.10$-s result is sensitive to the smoothing applied to the tracks (Fig.~\ref{FigMcKee}\textit{D}; Fig.~\ref*{SI-fig:si:empiricaltiming}C; \ref*{SI-SI:empiricaltiming}). Forward-response controls likewise show that longer spans can inflate fitted alignment (\ref*{SI-SI:test2}). The fish therefore carry a measurable heading contribution whose detection depends on the response definition, and the controls above show why this does not identify direct heading matching.

Walking pigeons \cite{naik2023pop,naik2023edmond} carry motion-capture markers, so their heading, measured as body yaw, is independent of travel. Social excess is resolved in dyads and in groups of five but not of ten, and no group size shows a heading excess with either centred or forward responses (Fig.~\ref{FigMcKee}\textit{C}; \ref*{SI-SI:pigeon:results}, \ref*{SI-SI:empiricaltiming}). The same birds expose the role of the observed channel. On identical observations from five-pigeon groups, replacing body yaw by travel direction, which also changes the turning response being fitted, raises the descriptive heading share of the social variance from $0.055$ to $0.218$ while the absolute heading increment falls (Fig.~\ref*{SI-fig:si:empiricaltiming}E,F). The share rises because the total social increment decreases, and the comparison does not decide which orientation the animals themselves use (\ref*{SI-SI:pigeon:results}). In 25-Hz recordings of tetras \cite{xue2023data,xue2023light}, whose head--tail axis is tracked from landmarks but whose front--back polarity is assigned from the direction of travel, reassigning neighbours across trials removes both increments, so the fitted structure is neighbour-specific, but this cross-trial null cannot separate direct heading matching from the artefacts above (\ref*{SI-SI:rn}, \ref*{SI-SI:cross}; Fig.~\ref*{SI-EDFig7}). Two further datasets serve narrower purposes. Tetras in an annular tank show that the transient disorder of collective U-turns makes bearing and heading easier to separate (\ref*{SI-SI:lech}), and larval zebrafish steered by an imposed light cue show the bearing sector in isolation, turning with a dominant first harmonic of the cue direction (\ref*{SI-SI:karpenko}).

\subsection*{Resolved heading-dependent structure departs from Vicsek form and from a common potential in the tracked coordinates}
\label{ResultsStructure}
\begin{figure}[tbp]
\centering
\includegraphics[width=\textwidth]{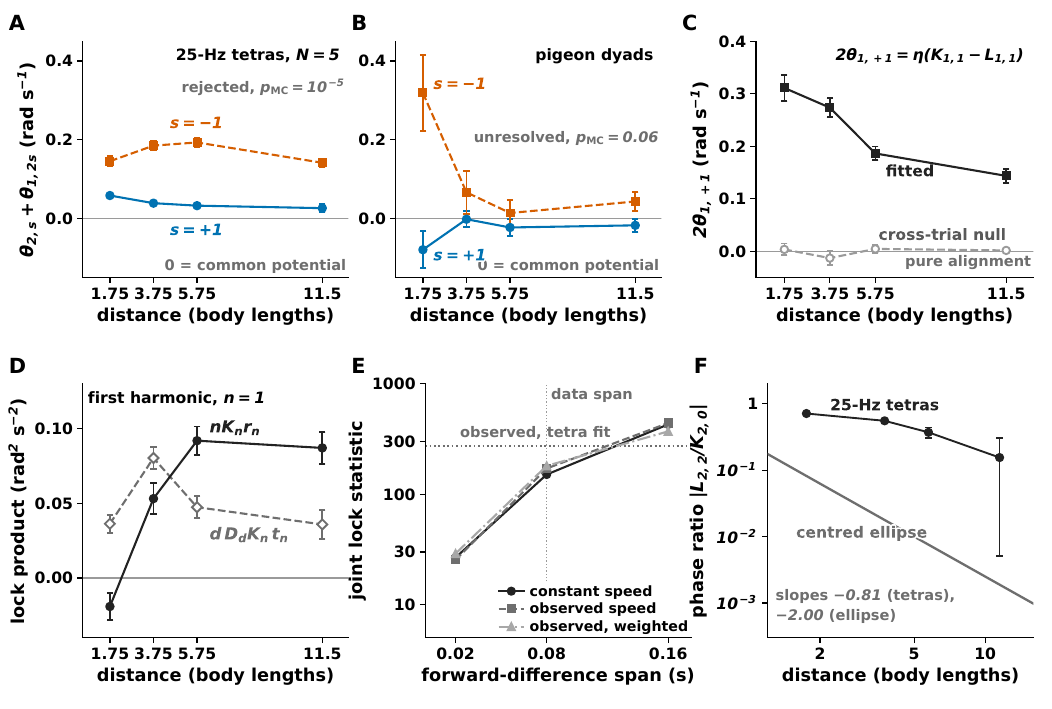}
\caption{\textbf{Coefficient relations test specific restrictions on the mechanism.} (\textit{A,B}) A common angular potential requires $\theta_{2,s}+\theta_{1,2s}=0$ for $s=+1$ (blue) and $s=-1$ (orange) at every distance. Five-fish groups of 25-Hz tetras \cite{xue2023data,xue2023light} (\textit{A}; bars one trial-jackknife SE over 26 trials) reject the restriction; pigeon dyads \cite{naik2023pop,naik2023edmond} (\textit{B}; bars one SE over 22 sequences) leave it unresolved; each panel reports the joint-test result ($p_{\mathrm{MC}}$, the restricted-bootstrap tail probability). (\textit{C}) In the same tetra fit, pure heading matching requires $\eta(K_{1,1}-L_{1,1})=2\theta_{1,+1}$ to vanish (the line marked pure alignment). The fitted component (squares) stays positive, while the cross-trial null (open circles) stays indistinguishable from zero; bars show one trial-jackknife SE. (\textit{D}) The coefficient lock requires the two plotted products to coincide, shown for $n=1$ from the speed-weighted lock fit, with $r_n=\theta_{n,-1}+\theta_{n,+1}$ and $t_n=\theta_{n,-1}-\theta_{n,+1}$ the radial and transverse channels, $K_n$ standing for $\theta_{n,0}$ and $D_dK_n$ the four-bin secant derivative; bars are one SE propagated from the trial-jackknife covariance. (\textit{E}) A continuous-time model that obeys the lock still gives a nonzero joint lock statistic after measurement over finite spans (log scale), for three speed treatments; the horizontal dotted line marks the observed joint statistic of the speed-weighted tetra fit (both harmonics; \textit{D} shows $n=1$) and the vertical dotted line the 0.08-s span of the data. Simulation and empirical covariance constructions differ, so this comparison does not quantify the fraction of the empirical discrepancy due to measurement. (\textit{F}) The fitted phase ratio $|L_{2,2}/K_{2,0}|$ of the same 25-Hz tetras (circles; bars one delta-method SE from the full covariance of 26 independent-trial jackknife fits) against the prediction of a uniform centred ellipse with $b/L=0.20$ (line; the predictions for $b/L$ from $0.15$ to $0.25$ lie within the line width); the fitted log--log slope is $-0.81$ against $-2$, and the comparison is descriptive. All panels are conditional on the tracked landmarks, fitted design and observed orientation, with front--back polarity assigned from motion in the tetra data. The even basis functions in \textit{F} are invariant to persistent polarity reversals, but their jointly fitted coefficients need not be (\ref*{SI-SI:forward:caveats}). \textit{A,B}: \ref*{SI-SI:forward:gradflow}, \ref*{SI-SI:pigeon:gradflow}; \textit{C}: \ref*{SI-SI:forward:vicsek}; \textit{D,E} and the second harmonic: \ref*{SI-SI:forward:lock}--\ref*{SI-SI:fwd:trialcluster}, Figs.~\ref*{SI-fig:si:inference}, \ref*{SI-fig:si:secondlock}; \textit{F}: \ref*{SI-SI:centro:null}, \ref*{SI-SI:forward:centro}.}
\label{FigStructural}
\end{figure}

Where a heading-dependent channel is resolved, the matrix specifies testable constraints. In the fitted coordinates a negative $m$ carries $\eta(K_{n,m}+L_{n,m})/2$ and a positive $m$ carries $\eta(K_{n,m}-L_{n,m})/2$, so that Eq.~\eqref{eq:mtintegrability} reads $\theta_{2,s}+\theta_{1,2s}=0$ for $s=\pm1$ (Materials and Methods). The 25-Hz tetra data are large enough to fit both signs of $m$ and the transposed interior cells, and their coefficients violate this condition (Fig.~\ref{FigStructural}\textit{A}): a restricted bootstrap over whole trials gives no exceedance in $99{,}999$ draws, a plus-one Monte Carlo probability $p_{\mathrm{MC}}=10^{-5}$, while simulated groups with an exact common potential measured over the same spans give statistics several times smaller (\ref*{SI-SI:forward:gradflow}, \ref*{SI-SI:fwd:trialcluster}). In pigeon dyads, where only 22 sequences support the bootstrap, the joint test is unresolved despite a visible near-field departure (Fig.~\ref{FigStructural}\textit{B}); separately, the dyad coefficients reject two specified bearing-only forward models with block covariance but not with sequence covariance (\ref*{SI-SI:pigeon:gradflow}; Fig.~\ref*{SI-fig:si:calibratednulls}d). The tetra $(1,1)$ cell is not of Vicsek form either. Pure heading matching requires the component that locks the sum of the two headings to the line of sight to vanish, and that component is positive at every distance while the cross-trial null is indistinguishable from zero (Fig.~\ref{FigStructural}\textit{C}; joint $p=1.1\times10^{-12}$; \ref*{SI-SI:forward:vicsek}, \ref*{SI-SI:fwd:trialcluster}).

The coefficient lock of Eq.~\eqref{eq:mtlock} is rejected as an exact relation among the binned bearing, radial and transverse coefficients in the four distance bins of Fig.~\ref{FigStructural}\textit{D}, with no exceedance in $9{,}999$ restricted draws, $p_{\mathrm{MC}}=10^{-4}$ (\ref*{SI-SI:forward:lock}--\ref*{SI-SI:fwd:trialcluster}). This does not reject perceptual delay. A continuous-time model that obeys the lock exactly, measured over the dataset's $0.08$-s response span, yields a large fitted violation (Fig.~\ref{FigStructural}\textit{E}), because finite temporal aggregation and the secant estimate of the radial derivative move the fitted coefficients off the discrete relation (\ref*{SI-SI:forward:lock}), whereas omitting the speed weight accounts for little of the observed statistic (\ref*{SI-SI:fwd:speed}; Fig.~\ref*{SI-fig:si:inference}). The departures from Vicsek form and from a common potential are statistically resolved in the tracked coordinates. They do not identify whether the violation arises from behaviour, from measurement or from misspecification of the fitted design, and they pool five illumination levels (\ref*{SI-SI:forward:gradflow}, \ref*{SI-SI:fwd:trialcluster}). Attributing them to the animals' rule rather than to the motion-assigned front--back polarity requires a polarity fixed without reference to motion in the same recordings, which these data do not provide (\ref*{SI-SI:fwd:resolver}, \ref*{SI-SI:forward:caveats}). One comparison uses even $(2,\pm2)$ and $(2,0)$ basis functions, which are unchanged by a persistent front--back reversal, although their jointly fitted coefficients can still depend on the other fitted terms. The point estimates of $|L_{2,2}/K_{2,0}|$ lie above the bin-centre prediction of a uniform centred ellipse and decay more slowly with distance, a descriptive comparison with substantial trial-level uncertainty at the farthest bin (Fig.~\ref{FigStructural}\textit{F}; \ref*{SI-SI:centro:null}, \ref*{SI-SI:forward:centro}, \ref*{SI-SI:fwd:resolver}, \ref*{SI-SI:forward:caveats}).

\section*{Discussion}
We asked whether social interaction in moving animal groups is organized by where neighbours \emph{are} or by where they are \emph{going}. The two descriptions are related by a common harmonic expansion. The bearing tradition and the alignment tradition are the edge and the diagonal of one coefficient matrix, indexed by the perceiver's bearing harmonic and the neighbour's aspect harmonic, and the image of an oriented neighbour supplies coefficients set by its geometry and the perceptual filters. An alignment-shaped coefficient then has at least three origins, each with its own signature: the projected image of an oriented neighbour, perceptual delay acting on a bearing response, and a direct response to heading.

The delay law makes the second origin concrete. Once the bearing profile is measured, its derivative fixes the distance dependence and the sign of the induced alignment coefficient, and the sign changes where $dK_1(d)$ is stationary rather than where attraction turns into repulsion. Part of the alignment reported from trajectories can be the shadow that a measured attraction profile casts through an animal's perceptual delay; how large that part is remains open. The coefficient lock tests this route without a separately estimated delay, with each neighbour's radial and transverse contributions weighted by its measured speed. A suitably matched optic-flow response obeys the same lock at first order, so manipulating the delay at fixed flow geometry is what would separate the two (\ref*{SI-SI:latencylock}, \ref*{SI-SI:opticflow}).

Comparing the two directed interactions answers a different question. Interactions that keep two equal headings equal are the even cells, and the diagonal is always among them, but preserving equality is not convergence. The stronger transpose relation decides whether both animals descend one angular potential at fixed positions. The point-stimulus theory does. Resolving neighbour orientation can remove this property, although each animal still follows its own landscape. Non-reciprocity of the directed landscape on the bearing edge coexists with a potential; only an interior violation creates curl. This connects a perceptual construction to the physics of non-reciprocal interactions \cite{Fruchart2021,Amichay2024}. The potential constrains angular relaxation at frozen positions only. Self-propelled motion changes the positions and can increase that potential, so no collective pattern alone can diagnose its absence (\ref*{SI-SI:integrability}).

The animal data show why a coefficient cannot be read independently of its measurement. Bearing responses are recovered in the 500-Hz tetras and social responses in the smaller pigeon groups. In the fish a heading excess appears with forward but not with centred turning responses, a dependence on the response definition that persists even when body orientation is measured independently of travel (\ref*{SI-SI:empiricaltiming}). Heading shares vary with the observed channel (Fig.~\ref*{SI-fig:si:empiricaltiming}E,F) and, in controls with known interactions, with the response span (\ref*{SI-SI:test2}), and those controls also show that a heading torque can go undetected while a bearing-only group can produce a positive heading excess. Non-detection is therefore neither an equivalence result nor evidence that heading matching is absent. The resolved tetra structure also departs from Vicsek form and from a common potential in the tracked coordinates, and the coefficient lock, tested on binned coefficients, is rejected without rejecting delay (\ref*{SI-SI:detectioncalibration}, \ref*{SI-SI:forward:caveats}).

These limits specify the requirements for future experiments. An imposed neighbour can vary body orientation at fixed bearing, or angular size at fixed image midpoint, and so remove the collinearity that polarized groups impose. A trajectory study needs body orientation from landmarks, a response span validated for the chosen response definition, independent biological replicates, enough configurations away from parallel motion, and positive controls run through the complete measurement procedure (\ref*{SI-SI:fom}, \ref*{SI-SI:detectioncalibration}, \ref*{SI-SI:cross}).

An inferred alignment coefficient thus carries a signature. Apparent size and image phase select the available sectors, the exchange relations constrain consensus and common potentials, and delay ties apparent alignment to the attraction profile. By connecting what an animal sees to how it turns, the bearing--heading spectrum offers a common frame for how individual perception becomes collective motion.

\section*{Materials and Methods}
\label{Methods}
\paragraph{Image geometry and the object filter.}
For an ellipse of length $L$ and width $b$ seen at aspect $\chi=\pb-\Tab$ and range $d$, the projected width is $w(\chi)=\sqrt{L^2\sin^2\chi+b^2\cos^2\chi}$ and the exact subtended angle is
\begin{equation*}
\Theta(\chi,d)=\operatorname{atan2}\Big(\sqrt{d^2w^2-L^2b^2/4},\;d^2-\tfrac{L^2+b^2}{4}\Big),
\end{equation*}
which tends to $w/d$ in the far field (\ref*{SI-SI:geometry}, \ref*{SI-SI:subtense}). The harmonic theory projects a stimulus onto the ring, smooths it with a sensory kernel of width $\sigma$ and overlaps it with the decision bump. The overlap is linear in the image, so an image of angular density $p$ contributes its characteristic function at frequency $n$ (\ref*{SI-SI:cumulant}). For a uniform silhouette the integral can be evaluated in elementary functions and gives Eq.~\eqref{eq:mtimage}. The image midpoint $\delta$ splits the signed amplitude into $K_n=Q_n\cos(n\delta)$ and $L_n=Q_n\sin(n\delta)$, and expanding $\Theta(\chi)$ and $\delta(\chi)$ in aspect harmonics gives Eq.~\eqref{eq:mttwoindex} (\ref*{SI-SI:objectfilter}--\ref*{SI-SI:aspectprofile}). For a centred ellipse, $\delta=\tfrac12\arg[1-((L^2-b^2)/4d^2)\,e^{2i\chi}]$, which is $\pi$-periodic and odd in $\chi$, so a centrosymmetric body has only even $m$ in both families (\ref*{SI-SI:centro}). All simulations and fits use the per-agent torque $\Gamma^{(a)}=-\eta\,\partial_{\pa}H^{(a\leftarrow b)}$. Differentiating the sum of the two directed landscapes defines a different model, which coincides with it only at $m=0$ (\ref*{SI-SI:consensus}, \ref*{SI-SI:prescriptions}).

\paragraph{Exchange and curl.}
With $u=\pa-\Tab$ and $w=\pb-\Tab$, exchanging the animals is the involution $(u,w)\mapsto(w-\pi,\,u-\pi)$, under which $e^{i(nu+mw)}\mapsto(-1)^{n+m}e^{i(mu+nw)}$, so a coefficient array maps to its parity-weighted transpose (\ref*{SI-SI:exchange}, \ref*{SI-SI:convention}). On the equal-heading manifold $u=w=x$ the relative torque of one cell is $R(x)=\eta n[1-(-1)^{n+m}][L_{n,m}\cos(nx)\sin(mx)-K_{n,m}\sin(nx)\cos(mx)]$, which vanishes for even $n+m$ (\ref*{SI-SI:consensus}). Both torques derive from one periodic potential if and only if $\partial_w\Gamma^{(a)}=\partial_u\Gamma^{(b)}$ at fixed landmarks. Substituting Eq.~\eqref{eq:mttwoindex} gives Eq.~\eqref{eq:mtintegrability}, in which the two families contribute independently and the factor $nm$ removes the edges (\ref*{SI-SI:integrability}). In fitted coordinates the condition reads $\theta_{m,n}=(-1)^{n+m}\theta_{n,m}$.

\paragraph{Perceptual delay.}
Let the perceived relative position relax to the true one with time constant $\tau$, so that $\hat d=d-\tau\dot d$ and $\hat T=T-\tau\dot T$ to first order. Inserting these into the landscape, differentiating with respect to heading, and only then substituting the pair kinematics for motion along body heading $\dot d=-(v_a\cos\psi_a+v_b\cos\psi_b)$ and $\dot T=-(v_a\sin\psi_a+v_b\sin\psi_b)/d$, with $\psi_b=\pb-\Tba$, gives
\begin{align}
\Gamma^{(a)}=-\eta\sum_{n\ge1}n\big\{&K_n\sin(n\psi_a)\nonumber\\
 +\tau\big[&nK_n\dot T\cos(n\psi_a)-K'_n\dot d\sin(n\psi_a)\big]\big\},
\label{eq:mtmaster}
\end{align}
valid when $n_{\max}\tau v/d\ll1$ and $\tau v/\xi\ll1$ for a typical speed $v$ and radial scale $\xi$ (\ref*{SI-SI:sensorydelay}--\ref*{SI-SI:differentiate}). The terms carrying $v_b$ form the pair torque, with a transverse part proportional to $nK_n d^{-1}\cos(n\psi_a)\sin\psi_b$ and a radial part proportional to $K'_n\sin(n\psi_a)\cos\psi_b$. Averaging over bearing at fixed heading difference removes every order but $n=1$ and gives Eq.~\eqref{eq:mtgalign}; the ratio of the two parts is Eq.~\eqref{eq:mtlock} (\ref*{SI-SI:neighbourmotion}, \ref*{SI-SI:latencylock}). The terms carrying $v_a$ are focal-motion corrections at orders $n\pm1$, and substituting the kinematics before differentiating gives them the wrong weights (\ref*{SI-SI:selfterms}). The lag is in the exteroceptive estimate of position; a lag of the animal's own decision bump is a different model.

\paragraph{Simulations.}
Behavioural simulations integrate the per-agent torque with rotational noise, constant speed, a point stimulus or the exact image, and body-scale positional exclusion that leaves headings unchanged. Matched comparisons share initial positions, headings and noise seeds and equate the mean first-harmonic gain at $d=\xi$. Image shape, intensity, harmonics, range, delay, noise and all parameters are given in \ref*{SI-SI:simulator} and \ref*{SI-SI:motion_atlas}, and the movies are described in \ref*{SI-SI:videos}.

\paragraph{Turning regression.}
Differentiating Eq.~\eqref{eq:mttwoindex} turns each landscape cell into signed predictors. For observation $i=(a,t)$ and distance interval $B_k$,
\begin{equation}
X_{i,n,m,k}=-n\sum_{b\ne a}\mathbf 1\{d_{ab,i}\in B_k\}\sin(n\psi_{ab,i}+m\chi_{ab,i}),
\label{eq:designcolumn}
\end{equation}
summed over neighbours before fitting, and the response is $y_i=\operatorname{wrap}[\phi_a(t+T_+)-\phi_a(t-T_-)]/(T_++T_-)$, centred when $T_+=T_-$ and forward when $T_-=0$. The regression $y=X\theta+Z\gamma+\epsilon$, with $Z$ holding the intercept and arena predictors, is fitted by weighted least squares without a ridge penalty \cite{Hoerl1970Ridge}; recorded-data weights are constant within a fit, so the estimates are ordinary least squares. The base design $\{(1,0),(2,0),(3,0),(1,-1),(2,-2)\}$ separates three bearing harmonics from polar and nematic alignment. Structural fits add both signs of $m$ and the transposed interior cells $(2,\pm1)$ and $(1,\pm2)$ in four distance bins. Fitted and landscape coefficients are related by $\theta_{n,\mp m}=\eta(K_{n,m}\pm L_{n,m})/2$ for $m>0$ and $\theta_{n,0}=\eta K_{n,0}$, so $(n,-n)$ is pure heading difference, pure Vicsek form in $(1,1)$ is $\theta_{1,+1}=0$, and the radial and transverse delay channels are $\theta_{n,-1}+\theta_{n,+1}$ and $\theta_{n,-1}-\theta_{n,+1}$. The fits in Fig.~\ref{FigMcKee}\textit{A--C} use centred responses, with matched forward refits in panel~\textit{D} and \ref*{SI-SI:empiricaltiming}; the tetra structural fits use forward increments, because centred differencing mixes the $(n,\pm1)$ channels at $25$~Hz (\ref*{SI-SI:pairdesign}, \ref*{SI-SI:twosector}, \ref*{SI-SI:forward:scheme}, \ref*{SI-SI:pigeon:spec}).

\paragraph{Null comparisons and inference.}
Let $X_b$ and $X_h$ be the bearing and heading columns of the design and $Z$ the arena block. The social and heading increments are $\Delta R^2_{\mathrm{soc}}=R^2(X_b,X_h,Z)-R^2(Z)$ and $\Delta R^2_{\mathrm{head}}=R^2(X_b,X_h,Z)-R^2(X_b,Z)$, and the tested quantity is the paired excess $E_k=\Delta R^2_{k,\mathrm{real}}-\Delta R^2_{k,\mathrm{null}}$ for $k\in\{\mathrm{soc},\mathrm{head}\}$ over a matched null on the same eligible observations; $E_{\mathrm{head}}$ is the heading excess. For the 500-Hz tetras the null assigns each focal fish the neighbours of a donor trial, from the same recording day when one exists, cycled to the focal length; for the pigeons it assigns a donor segment from another sequence of at least the focal length. In the 25-Hz tetras a cross-trial neighbour null serves as a separate diagnostic. Recording day is the independent unit for the 500-Hz tetras, sequence for the pigeons and trial for the 25-Hz tetras. Delete-one-day jackknives for the 500-Hz tetras, sequence-rematching bootstraps for the pigeons and, for the structural restrictions, covariances clustered by trial (25-Hz tetras) or sequence (pigeon dyads) with restricted wild-cluster bootstraps over those units that impose each null before resampling \cite{Cameron2008Bootstrap,Webb2023} supply the intervals and tail probabilities (\ref*{SI-SI:twosector}--\ref*{SI-SI:dayunit}, \ref*{SI-SI:rn:nulls}, \ref*{SI-SI:fwd:trialcluster}, \ref*{SI-SI:pigeon:spec}--\ref*{SI-SI:pigeon:gradflow}).

\paragraph{Positive controls.}
We added known torques $A\sin[n(\phi_b-\phi_a)]$ per neighbour, for which $\theta_{n,-n}=A/n$, to bearing-only dynamics whose drift is the bearing, wall and arena response fitted to each dataset with the heading terms removed. Positions were either held at their recorded values or moved with the simulated headings, and observation error was added or omitted. Each repetition runs the preprocessing, null construction and clustered inference of the real analysis. Recovery of the coefficient, detection of a positive heading excess and coverage of the heading excess that the generator itself yields after observation and fitting are separate targets. Matched fish controls compare both response definitions under zero-phase filtering, one-sided filtering with the same magnitude response and no filtering (\ref*{SI-SI:detectioncalibration}, \ref*{SI-SI:filtertiming}).

\paragraph{Data and code.}
The analyses use five public deposits: 500-Hz schooling tetras \cite{mckee2020,mchenry2020dryad}, 25-Hz tetras under varied illumination \cite{xue2023data,xue2023light}, tetras in an annular tank \cite{Lecheval2018,lecheval2018dryad}, marker-tracked pigeons \cite{naik2023pop,naik2023edmond} and larval zebrafish phototaxis \cite{Karpenko2020,karpenko2020dryad}. Their versions, units, trial selection and preprocessing are given in \ref*{SI-SI:data}--\ref*{SI-SI:karpenko}. The 500-Hz recordings were also analysed in the harmonic theory paper for temporal spectra and collective statistics \cite{Salahshour2025Harmonic}; here they serve to estimate bearing and heading responses. Analysis and simulation code, and supplementary movies are available in Figshare \cite{Salahshour2026BearingHeadingSoftware}.

\paragraph{AI assistance.}
Anthropic Claude (Opus 5, Opus 5.5, Fable 5 and Fable 5.1) and OpenAI ChatGPT/Codex (5.6 Sol and Astra 6) were used for language edits and readability suggestions, programming, mathematical and numerical analysis, algebraic and consistency checks, plotting code, dataset and literature searches, and manuscript review. The author specified the analyses, checked the derivations, reviewed and verified the outputs used in this study, and takes full responsibility for the content of the paper.

\Needspace{8\baselineskip}
\section*{Acknowledgement}
The author thanks Iain D. Couzin for his generous support and insightful discussions, and the researchers---Amberle McKee, Matthew J. McHenry, Tingting Xue, Valentin Lecheval, Hemal Naik, Sophia Karpenko, together with their coauthors---whose experimental work and publicly shared datasets made these analyses possible. This work was supported by the Deutsche Forschungsgemeinschaft (German Research Foundation) under Germany's Excellence Strategy---EXC 2117--422037984.

\bibliographystyle{unsrtnat}
\bibliography{refs}
\end{document}